\documentclass[11pt]{iopart}
\usepackage{iopams}
\usepackage{epsfig}
\usepackage{amsfonts,amsthm,eucal,amssymb}
\usepackage{color}
\usepackage{graphicx}
\eqnobysec
\begin{document}

\jl{3}

\title[Dirac-Weyl equation on a hyperbolic graphene surface under magnetic fields]{Dirac-Weyl equation on a hyperbolic graphene surface under perpendicular magnetic fields}
\author{
D Demir K{\i}z{\i}l{\i}rmak${}^a$, \c{S} Kuru${}^b$, J
Negro${}^c$}

\address{${}^a$Department of Medical Services and Techniques, Ankara Medipol University, 06050 Ankara, Turkey\\
${}^b$Department of Physics, Faculty of Science, Ankara
University, 06100 Ankara, Turkey\\
${}^c$Departamento de F\'{\i}sica Te\'orica, At\'omica y
\'Optica, Universidad de Valladolid, 47071 Valladolid, Spain }

\ead{duygudemirkizilirmak@gmail.com,kuru@science.ankara.edu.tr,
jnegro@fta.uva.es}

\bigskip
\bigskip
\begin{flushright}
\small{{\it This paper is dedicated to the memory of our dear colleague \\ {\bf Georgy Pavlovich Pronko}, who passed away on October 27, 2019.}}
\end{flushright}
\bigskip

\begin{abstract}
In this paper the Dirac-Weyl equation on a hyperbolic surface of graphene under magnetic fields is considered. In order to solve this equation analytically for some cases, we will deal with vector potentials symmetric under rotations around the $z$ axis. Instead of using tetrads we will get this equation from a more intuitive point of view by restriction from the Dirac-Weyl equation of an ambient space. The eigenvalues and  corresponding eigenfunctions for some magnetic fields are found by means of the factorization method. The existence of a zero energy ground
level and its degeneracy is also analysed in relation to the Aharonov-Casher theorem valid for flat graphene. 
\end{abstract}




\noindent



\section{Introduction}

The low energy electrons in flat graphene behave in the continuum limit as massless Dirac particles. Based on this property, there has been a considerable amount of work on the electronic properties of graphene and other allotropes of carbon under different magnetic or electric fields by making use of the (2+1) dimensional Dirac-Weyl equation  \cite{Semenoff,Novoselov1,Peres,Kuru1,Goerbig,Downing,Hartmann}. 
Another attractive field of research has been the study of Dirac electrons in non-flat surfaces specially fullerenes (or nanotubes) which can be addressed by expressing the Dirac-Weyl equation on the sphere (or on the cylinder by means of appropriate boundary conditions) and  nano-ribbons \cite{Vit1,Roy1,Vit2,Lee,Pnueli,Fakhri,Abrikosov,Pudlak,Brey,Gorbar}. 

In the same way, the electronic properties of massless Dirac electrons in a graphene surface with hyperbolic shape, together with the presence of external electromagnetic fields, can also be studied by means of  the Dirac-Weyl equation on this surface. This is the main objective of the present paper, where we will consider only perpendicular magnetic fields with rotational symmetry around the $z$-axis. Usually the construction of the Dirac-Weyl equation on a curved surface is obtained with the help of covariant derivatives with spin connections \cite{Fakhri,Abrikosov,Pudlak}. Here, we will adopt a simpler approach by means of the restriction from the standard Dirac equation defined in an ambient space to a surface included in this space.
This point of view is easier to follow and, in particular for constant curvature
surfaces, it allows to keep track of the explicit symmetries of the surface. We remark that this approach leads to equivalent results as those
obtained in the usual formulation as it has been checked with the case of the sphere \cite{Vit1,Roy1,Roy2}.

An application of this study will be the finding of the energy levels of
the Landau system on the graphene hyperboloid as well as their degeneracy.
In this respect, we will see that the number of energy levels is finite, each one with infinite degeneracy.

This type of surface can be seen as taking part of a quantum blister. In bilayer graphene some patchs are deformed and they lead to quantum blisters \cite{Abdullah1,Abdullah2}. By applying magnetic fields in the hyperbolic surface we can study the confining of Dirac electrons on this type of deformations.
Recently, it has been shown that electrons can be confined also on quantum blisters by applying  electrostatic voltage
\cite{Liu}.

Graphene can be deformed to produce surfaces with curvature (as the hyperboloid of this paper) by means of many other different techniques. Very recently, graphene bubbles with many shapes have been obtained by using the tip in the atomic force microscopy (AFM) \cite{Jia19}. Graphene bubbles have been detected by scanning tunnelling microscopy and pseudo Landau levels associated to strain of the lattice have been observed by STS spectra \cite{Levy10}. The shapes of graphene nano-bubbles can also be modified by means of external electric fields \cite{Georgiou11}. This type of deformation under the action of external magnetic fields would be a natural set up to realize the experiments on our system.
Other kind of deformations of graphene has been obtained as defects in the honey-comb lattice by including pentagons or heptagons that give rise to curvature on the flat graphene leading to the form of cones or to (one and two) sheeted hyperboloids \cite{Fukuyama01, Vozmedian07, Smotlacha11}.

The graphene lattice is subject to strain, for example from the shape of bubbles, or due to defects of heptagones or pentagons inserted in the honey-comb periodic structure producing some kind of curvature. The effect of strains on the hopping parameters is the origin of changes in the electronic structure which are described by means of gauge fields giving rise to a type of pseudo magnetic fields \cite{Pincak03, Kolesnikov04, Guinea10N}. In this paper we have restricted ourselves to the influence of external magnetic fields on the electronic states of graphene having the form of a hyperbolic two sheeted manifold. The aim is to get analytical expressions for some external symmetric magnetic fields. Based on these results our plan in the near future is to find the qualitative changes produced by the pseudo magnetic fields as perturbation of the external `true' magnetic fields.

The organization of this paper is as follows. In Section 2, the Dirac-Weyl equation on the hyperboloid  is defined. In Section 3, the factorization method of supersymmetric quantum mechanics is introduced in order to solve this equation and the ground state solutions are characterized. The relation between the existence of good ground states and the  magnetic flux is discussed
taking as reference the Aharonov-Casher theorem valid for magnetic fields acting on flat graphene. Next, a few solvable cases are worked out in Section 4. Finally, this work is finished with some conclusions and remarks along Section 5.

\section{The Dirac-Weyl equation on the hyperboloid}

Low energy electrons in graphene behave as massless Dirac electrons with an
effective  Fermi velocity $\upsilon_F=c/300$, where $c$ is the
velocity of light  (see for instance \cite{Semenoff}). 
Therefore, they are described by the $2{+}1$
dimensional Dirac-Weyl equation in flat space-time. This description
can be extended to other surfaces, in particular there are
many recent works devoted to adapt it to fullerenes and nanotubes.
Here, we will study the Dirac-Weyl equation on the two--dimensional hyperboloid. 
Our method consists in formulating the Dirac-Weyl equation in an
ambient space, where the spatial components $(x,y,z)$ have metric signature
$(-,-,+)$. In this space we will restrict the Dirac-Weyl equation
to the hyperboloid $-x^2-y^2+z^2=c$, where $c= r^2$ is the square of the `radius' 
of the sheet $z>0$ of a two--sheeted hyperboloid. In this way, we will
get a Dirac-Weyl equation on the hyperboloid which inherits the
$SO(1,2)$ symmetry valid on the whole ambient space.

\subsection{Reduction of Dirac-Weyl equation to the hyperboloid}

The Dirac-Weyl equation in $3{+}1$ space-time for Cartesian coordinates is given by
\begin{equation}\label{detd}
i\hbar\,\frac{\partial
\Phi(x,y,z,t)}{\partial t} = \upsilon_F\,({\pmb\sigma}\cdot\mathbf{p})\Phi(x,y,z,t) \, ,
\end{equation}
where $\pmb{\sigma}=(\sigma_x,\sigma_y,\sigma_z)$ are the Pauli
matrices and $\mathbf{p}=-i\hbar\,
(\partial_x,\partial_y,\partial_z)$ is the three dimensional
momentum operator. The interaction of a Dirac electron with a
magnetic field according to the minimal coupling rule is described
by  replacing the momentum operator $\mathbf{p}$ in (\ref{detd}) by
$\mathbf{p}-q\mathbf{A}/c$, where the charge of the electron is
$q=-e$. The notation for the vector potential and magnetic field is the
usual one
\begin{equation}\label{ab}
\mathbf{A}=(A_x,A_y,A_z),\qquad \mathbf{B}=\nabla\times\mathbf{A}\, .
\end{equation}
The time-independent Dirac-Weyl equation, obtained by replacing $\Phi(x,y,z,t)=\Psi(x,y,z)\,e^{-i E t/\hbar}$ into (\ref{detd}), is
\begin{equation}\label{dea}
\upsilon_F\,\left[{\pmb\sigma}\cdot(\mathbf{p}-\frac{q}c\,\mathbf{A})\right]\Psi(x,y,z)=E\,\Psi(x,y,z)
\,.
\end{equation}

In the following, we will adapt the above equation (\ref{dea}) to our present situation.
Firstly, in order to keep the formal $SO(1,2)$ symmetry, we must use everywhere the scalar
product with signature $(-,-,+)$ represented by the dot $``\cdot"$ instead of the Euclidean inner product.

Let us introduce the hyperbolic (or pseudo--spheric) coordinates $(r,u,\varphi)$ appropriate to describe a two--sheeted
hyperboloid  oriented along the $z$ axis satisfying the equation 
$-x^2-y^2+z^2=r^2$. They are  related to
the Cartesian coordinates $(x,y,z)$ by 
\begin{equation}\label{s}
x=r\sinh u \cos\varphi,\quad y=r\sinh u \sin\varphi,\quad z=r\cosh
u\, ,
\end{equation}
where $0<u<\infty$, $0\leq\varphi<2\pi$ and $0<r<\infty$. The momentum
operators in hyperbolic coordinates are
\begin{equation}\label{px}
p_x=-i \hbar\,\partial_x=-i \hbar(-\sinh u
\cos\varphi\,\partial_r-\frac{\sin\varphi}{r\sinh
u}\,\partial_{\varphi}+\frac{\cosh u
\cos\varphi}{r}\,\partial_{u})\,,
\end{equation}
\begin{equation}\label{py}
p_y=-i \hbar\,\partial_y=-i \hbar(-\sinh u
\sin\varphi\,\partial_r+\frac{\cos\varphi}{r\sinh
u}\,\partial_{\varphi}+\frac{\cosh u
\sin\varphi}{r}\,\partial_{u})\,,
\end{equation}
\begin{equation}\label{pz}
p_z=-i \hbar\,\partial_z=-i \hbar(\cosh u \,\partial_r-\frac{\sinh u
}{r}\,\partial_{u})\,.
\end{equation}

In a second step, in (\ref{dea}) we must use not arbitrary momenta,
but those restricted to the tangent plane of the hyperboloid.
They are defined in the way shown in \cite{Gadella11,Gadella13}:
\begin{eqnarray}\label{pn}
\begin{array}{c}
\tilde {p}_x=\frac{1}{2}(r\,p_x+p_x\,r)+\frac{1}{2}\left((\mathbf{r}\cdot\mathbf{p})\frac{x}{r}+\frac{x}{r}(\mathbf{p}\cdot\mathbf{r})\right)\,,\\[1ex]
\tilde {p}_y=\frac{1}{2}(r\,p_y+p_y\,r)+\frac{1}{2}\left((\mathbf{r}\cdot\mathbf{p})\frac{y}{r}+\frac{y}{r}(\mathbf{p}\cdot\mathbf{r})\right)\,,\\[1ex]
\tilde
{p}_z=\frac{1}{2}(r\,p_z+p_z\,r)-\frac{1}{2}\left((\mathbf{r}\cdot\mathbf{p})\frac{z}{r}+\frac{z}{r}(\mathbf{p}\cdot\mathbf{r})\right)\,.
\end{array}
\end{eqnarray}
These operators  are the quantum analog of the projection (according to the pseudo--scalar product) of the
momentum vectors on the tangent plane at a point of the hyperboloid.
They can be identified as the angular momenta.
On the hyperbolic surface where $r=R=const.$ 
the linear momentum operators are given by dividing by the constant radius, thus leading to the following expressions
\begin{equation}\label{pnx}
\hat {p}_x=\frac{-i \hbar}{R}(-\frac{\sin\varphi}{\sinh
u}\,\partial_{\varphi}+\cosh u \cos\varphi\,\partial_{u}+\sinh u
\cos\varphi)\,,
\end{equation}
\begin{equation}\label{pny}
\hat {p}_y=\frac{-i \hbar}{R}(\frac{\cos\varphi}{\sinh
u}\,\partial_{\varphi}+\cosh u \sin\varphi\,\partial_{u}+\sinh u
\sin\varphi)\,,
\end{equation}
\begin{equation}\label{pnz}
\hat {p}_z=\frac{-i \hbar}{R}(-\sinh u \,\partial_{u}-\cosh u)\,.
\end{equation}

Finally, we have to use the Dirac matrices appropriate to the metric. In this case,
the time-space metric is $g^{\mu\nu} = {\rm diag}(1,-1,-1,1)$. A choice for the Pauli matrices is $\hat{\pmb\sigma}=(-\sigma_x,-\sigma_y,i \sigma_z):=(\hat \sigma_1,\hat \sigma_2,\hat \sigma_3)$, so that
\begin{equation}
\hat\sigma_k\hat\sigma_j+ \hat\sigma_j\hat\sigma_k= 0 \quad (k\neq j),\qquad
\hat\sigma_1^2=\hat\sigma_2^2=1,\  \hat\sigma_3^2=-1\,.
\end{equation}
The Dirac matrices in terms of the previous ones are constructed, in the Weyl representation, as
\begin{equation}
\gamma^0 =\left(\begin{array}{cc} 0 & I \\ I & 0 \end{array}\right),\quad
\gamma^i =\left(\begin{array}{cc} 0 & \hat \sigma_i \\ -\hat \sigma_i & 0 \end{array}\right)\,.
\end{equation}
For this  choice of the Dirac matrices the dot product in the stationary Dirac-Weyl equation in (\ref{dea}) should be replaced by
\begin{equation}\label{sigmap}
\hat{\pmb\sigma}\cdot\hat{\mathbf{p}}=-\sigma_x\,\hat
{p}_x-\sigma_y\,\hat {p}_y+i\sigma_z\,\hat {p}_z\,.
\end{equation}

We will consider a magnetic field perpendicular to the surface of
the hyperboloid and having a rotational symmetry around the
$z$-axis. Hence, we choose the vector potential in the form
\begin{equation}\label{a}
\mathbf{A}=A(u) \hat{\varphi}=A(u) (-\sin \varphi,\cos\varphi,0)\, ,
\end{equation}
where $A(u)$ is a function depending on $u$.

Using the above definitions, after straightforward computations, the Hamiltonian corresponding to the
Dirac electron (\ref{dea}) on the surface of the hyperboloid becomes
\begin{eqnarray}\nonumber
\fl 
H=\frac{1}{R}\,\left(
  \begin{array}{c}
    (-\sinh u \,\partial_{u}-\cosh u)\qquad
   i\,e^{-i\varphi}(-\frac{i}{\sinh u}\,\partial_{\varphi}+{\cosh u}\,\partial_{u}+\sinh u-\frac{q\,R}{c\,\hbar}A(u)) \\[2ex]
     i\,e^{i\varphi}(\frac{i}{\sinh u}\,\partial_{\varphi}+{\cosh u}\,\partial_{u}+\sinh u+\frac{q\,R}{c\,\hbar}A(u))
     \qquad - (-\sinh u \,\partial_{u}-\cosh u)
  \end{array}
\right)\\\label{hsc}
\end{eqnarray}
and the eigenvalue equation for $H$, after dividing by $\hbar\,\upsilon_F$, is
\begin{equation}\label{dwsc}
H\,\Psi(R,u,\varphi)={\cal{E}}\,\Psi(R,u,\varphi),\qquad
{\cal{E}}=\frac{E}{\hbar\,\upsilon_F}\, .
\end{equation}

\subsection{Rotational symmetry}

Since the radius $R$ is constant, the notation $\Psi(R,u,\varphi):=
\Psi(u,\varphi)$ will be used hereafter, where $(u,\varphi)$ are a kind of
polar coordinates on the hyperboloid. Next, let us consider the total angular
momentum along the $z$ axis,
\begin{equation}\label{tam}
J_z=-i\hbar \partial_{\varphi}+\frac{\hbar}{2}\sigma_z\, .
\end{equation}
As there is a geometric rotational symmetry around the $z$-axis, $J_z$ should commute with $H$:
$[H,J_z]=0$.
Therefore, the eigenfunctions (\ref{dwsc}) of $H$ can also
be chosen as eigenfunctions of $J_z$ at the same time,
\begin{equation}\label{j}
J_z\Psi(u,\varphi)=\lambda\,\hbar\,\Psi(u,\varphi)\, .
\end{equation}
So, the two-component spinor wavefunction takes the form
\begin{eqnarray}\label{ppt}
\Psi(u,\varphi)=N\,\left(
  \begin{array}{c}
   e^{i(\lambda-\frac{1}{2})\varphi}f_1(u)\\[1ex]
     e^{i(\lambda+\frac{1}{2})\varphi}f_2(u)\end{array}
\right)\, ,
\end{eqnarray}
where $\lambda$ is a half-odd number and $N$ is a normalization
constant. By substituting (\ref{ppt}) and (\ref{hsc}) into the eigenvalue 
equation
(\ref{dwsc}) we get
\begin{equation}
 \begin{array}{c}
\fl
\displaystyle\left[\frac{1}{R}(-\sinh u \,\partial_{u}-\cosh u)\sigma_z
+(\frac{i}{2\,R \sinh u}+\frac{i}{R}\cosh
u\,\partial_{u}+\frac{i}{R}\sinh u )\sigma_x\right.
\\ [2ex] 
\displaystyle\left. +(\frac{q}{c\,\hbar}A(u)-\frac{\lambda}{R\,\sinh u})\sigma_y \right]
F(u) ={\cal E} F(u)\, ,\\\label{dwsc1}
\end{array}
\end{equation}
where $F(u)=(f_1(u),f_2(u))^T$ is a column matrix; the superindex
$T$ is used for matrix transposition. 

\subsection{Hermitian form}

In order to eliminate the
term with $\sigma_z$, we apply a transformation to the matrix
equation (\ref{dwsc1})
\begin{eqnarray}\label{ft}
F(u)=e^{-\frac{u}{2} \sigma_y} \left(
  \begin{array}{c}
   \psi_1(u)\\[1ex]
     \psi_2(u)
     \end{array}
\right)\,.
\end{eqnarray}
Then,  after using  the
Baker-Campbell-Hausdorff formula, Eq.~(\ref{dwsc1}) becomes
\begin{eqnarray}\label{hsc1}
\fl
\left(
  \begin{array}{c}
   0\qquad \frac{i}{R}\partial_{u}+\frac{i}{2\,R}\coth{u}-i(\frac{q}{c\,\hbar}A(u)-\frac{\lambda}{R\,\sinh{u}}) \\[1.5ex]
   \frac{i}{R}\partial_{u}+\frac{i}{2\,R}\coth{u}+i(\frac{q}{c\,\hbar}A(u)-\frac{\lambda}{R\,\sinh{u}})\qquad 0  \end{array}
\right)\left(
  \begin{array}{c}
   \psi_1(u)\\[1ex]
     \psi_2(u)
     \end{array}
\right)={\cal E}\left(
  \begin{array}{c}
   \psi_1(u)\\[1.5ex]
     \psi_2(u)
     \end{array}
\right)\,.
\end{eqnarray}
This effective Hamiltonian (\ref{hsc1}) is not  Hermitian due to
the term ``$\frac{i}{2\,R}\coth{u}$".  But it can be made Hermitian by
writing the wavefunction $(\psi_1(u),\psi_2(u))^T$ as
\begin{equation}\label{gp1}
(\psi_1(u),\psi_2(u))^T=\frac{1}{\sqrt{\sinh{u}}}(g_1(u),
ig_2(u))^T\, = \frac{1}{\sqrt{\sinh{u}}}\, G(u).
\end{equation}
It is  clear that the latter change is related to the surface element 
$ds =R^2  \sinh u \, du d\varphi$ of the hyperboloid in polar coordinates.
Thus, after these transformations we arrive at an effective Hermitian matrix Hamiltonian that can be expressed as
\begin{eqnarray}\label{hsc2}
\left(
  \begin{array}{c}
   0\qquad \frac{i}{R}\partial_{u}+i(\frac{\lambda}{R\,\sinh{u}}-\frac{q}{c\,\hbar}A(u)) \\[1ex]
  \frac{i}{R}\partial_{u}-i(\frac{\lambda}{R\,\sinh{u}}-\frac{q}{c\,\hbar}A(u))\qquad 0  \end{array}
\right)\left(
  \begin{array}{c}
   g_1(u)\\[1.5ex]
    i g_2(u)
     \end{array}
\right)={\cal E}\left(
  \begin{array}{c}
   g_1(u)\\[1.5ex]
     i g_2(u)
     \end{array}
\right).
\end{eqnarray}

\section{SUSY partner Hamiltonians and ground states}

\subsection{The supersymmetry formalism}

Let us define the following first order operators
\begin{equation}\label{lpm}
L^{\pm}=\mp \partial_{u}+W(u), \qquad
W(u)=-\frac{\lambda}{\sinh{u}}+\frac{q\,R}{c\,\hbar}A(u) ,
\end{equation}
where $W(u)$ is called superpotential function. Using these
definitions, the matrix equation (\ref{hsc2}) is rewritten as
\begin{eqnarray}\label{hsc3}
\left(
  \begin{array}{c}
   0\qquad -i\,L^+ \\[1ex]
  i\,L^-\qquad 0  \end{array}
\right)\left(
  \begin{array}{c}
   g_1(u)\\[1ex]
    i g_2(u)
     \end{array}
\right)=R\,{\cal E}\left(
  \begin{array}{c}
   g_1(u)\\[1ex]
     i g_2(u)
     \end{array}
\right)
\end{eqnarray}
and the components $g_1,g_2$ are connected by these operators,
\begin{equation}\label{e12}
L^+ g_2(u)=R\,{\cal E}\,g_1(u)\,, \qquad L^-g_1(u)=R\,{\cal
E}\,g_2(u)\,   .
\end{equation}
From these equations we get a pair of decoupled second order effective Schr\"odinger 
equations
\begin{eqnarray}\label{h1}
H_1\,g_1(u):=L^+L^-\,g_1(u)=\varepsilon\,g_1(u)\,  ,
\\
\label{h2} H_2\,g_2(u):=L^-L^+\,g_2(u)=\varepsilon\,g_2(u)\, ,
\end{eqnarray}
where $\varepsilon=R^2{\cal E}^2$. Equations (\ref{h1}) and
(\ref{h2}) in matrix form are
\begin{eqnarray}\label{hsc4}
\left(
  \begin{array}{c}
   L^+L^-\qquad 0 \\[1ex]
  0\qquad L^-L^+  \end{array}
\right)\left(
  \begin{array}{c}
   g_1(u)\\[1ex]
    i g_2(u)
     \end{array}
\right)=\varepsilon \left(
  \begin{array}{c}
   g_1(u)\\[1ex]
     i g_2(u)
     \end{array}
\right) .
\end{eqnarray}
The diagonal elements of the above matrix are the effective Hamiltonians 
\begin{equation}\label{h12}
H_1=-\partial_{u}^2+V_1(u),\qquad H_2=-\partial_{u}^2+V_2(u),
\end{equation}
whose effective potentials are given in terms of the superpotential (\ref{lpm}) in the following way
\begin{equation}\label{v12}
V_1(u)=W(u)^2-W'(u),\qquad V_2(u)=W(u)^2+W'(u)\,.
\end{equation}
Here, the prime denotes differentiation with respect to $u$. The above
relations show that the Hamiltonians $H_1$ and $H_2$  are one
dimensional supersymmetric partner Hamiltonians \cite{Cooper} and
$L^{\pm}$ are intertwining operators that link these Hamiltonians as follows:
\begin{equation}\label{l12}
H_2\,L^-=L^-\,H_1,\qquad H_1\,L^+=L^+\,H_2 .
\end{equation}
These intertwining relations imply that if we assume that the spectrum of $H_1$ ($H_2$) is
known then its partner $H_2$ ($H_1$) will have the same spectrum except possibly  the ground state.

Let $\{ \varepsilon_{1,n} \}$, $n=0,1,\dots$, be the discrete
spectrum of $H_1$ with real eigenfunctions $\{ g_{1,n}
\}$, and assume that the ground state of $H_1$ is annihilated by
$L^-$,
\begin{equation}\label{lp2}
L^-\, g_{1,0} = 0 \,  .
\end{equation}
Then, as a consequence of (\ref{h1})  
the ground
state eigenvalue of $H_1$ will be
\[
\varepsilon_{1,0} =0 \, .
\]
This will be a `good' ground state as far as the function $g_{1,0}$
is square-integrable in $(0,\infty)$ and satisfies appropriate
boundary conditions. Now, the discrete spectrum of $H_2$ will
consist of the eigenvalues $\{ \varepsilon_{2,n-1}\}$ and normalized
eigenfunctions $\{ g_{2,n-1}\}$ given by
\begin{equation}\label{pe12b}
\varepsilon_{1,n} =\varepsilon_{2,n-1} , \qquad g_{2,n-1} (u):=
\frac{1}{\sqrt{\varepsilon_{1,n} }} \,L^-\,g_{1,n} (u), \qquad
n=1,2,\dots \,.
\end{equation}
In this point, it is assumed that the operator $L^-$ does not spoil the
physical requirements of the eigenfunctions.
Thus, the  eigenvalues of the equations (\ref{hsc4}) for $g_1$ and $g_2$ consistent with
(\ref{hsc3})  are
\[
\varepsilon_{0} :=\varepsilon_{1,0}=0,\qquad \varepsilon_{n}
:=\varepsilon_{1,n} =\varepsilon_{2,n-1},\qquad n=1,2,\dots\,.
\]
Taking into account the above results, the excited
eigenfunctions of the reduced Hamiltonian Eq.~(\ref{hsc2}) take the form
\begin{eqnarray}\label{png}
G_{\pm,n}(u)=N \left(
  \begin{array}{c}
  \pm 
  g_{1,n}(u)\\[1ex]
    i\, 
    g_{2,n-1}(u)\end{array}
\right)
\end{eqnarray}
with the corresponding eigenvalues
\begin{equation}\label{evg}
{\cal{E}}_{\pm,n}:=\frac{E}{\hbar\,\upsilon_F}=
\pm\frac{1}{R}\sqrt{\varepsilon_{n}},
\qquad  n=1,2,\dots\,.
\end{equation}
The ground state wavefunction and its energy are as follows
\begin{eqnarray}\label{pnev}
 G_0(u)=N
\left(
  \begin{array}{c}
  g_{1,0}(u)\\[1ex]
    0\end{array}
\right),\qquad{\cal{E}}_{\pm,0}=\frac{1}{R}\sqrt{\varepsilon_{0}}=0\, .
\end{eqnarray}

There are other possibilities to characterize the ground state besides (\ref{lp2}),
for instance
\begin{equation}\label{lp3}
L^+\, g_{2,0} = 0 \,  .
\end{equation}
Or even a ground state not satisfying (\ref{lp2}) nor (\ref{lp3}), however in
our examples the present assumption will be sufficient.

\subsection{The zero energy ground state and the flux of the magnetic field}

Let us pay attention to the zero energy ground state wavefunction $g_{1,0}$ defined by (\ref{lp2})
that according to (\ref{lpm}) is determined by the equation
\begin{equation}
 \left(\partial_{u}-\frac{\lambda}{\sinh{u}}+\frac{q\,R}{c\,\hbar}A(u)\right)
 g_{1,0}(u) = 0 ,
\end{equation}
whose solution is
\begin{equation}\label{g10}
g_{1,0}(u)= N \left(\tanh\frac{u}2\right)^\lambda e^{-\frac{q\,R}{c\,\hbar}\int A(u) du}\,.
\end{equation}
In order $g_{1,0}(u)$ to be a physical ground state it should have
the following asymptotic behaviour:
\begin{equation}\label{behaviour}
\begin{array}{lll}
(a)\quad {\rm in}\  &u \to 0,\quad &g_{1,0}(u) \to 0 \ {\rm (or\ be \ bounded)}\,,
\\[1.5ex]
(b)\quad {\rm in}\  &u \to \infty,\quad &g_{1,0}(u) \to 0\,.
\end{array}
\end{equation}
Next, we will interpret the integral in the exponent of (\ref{g10}). Let us recall that the magnetic field is given by
\begin{equation}\label{bb}
{B}_{u,\varphi}(u)=\frac{1}{R\,\sinh{u}}\left[\partial_{u}(A(u)\,\sinh{u})\right].\end{equation}
Then, the magnetic flux $\Phi(u)$ in the circle of radius $u$ will be
\begin{equation}
\Phi(u) = \int_0^u B_{u,\varphi}(u) 2\pi R^2 \sinh u\, du= 2\pi R A(u) \sinh u \,,
\end{equation}
therefore,
\begin{equation}
A(u) = \frac{\Phi(u)}{2\pi R \sinh u} \, .
\end{equation}
Now, assume that we have a null magnetic field for $u>u_0$, then in the
region $0<u<u_0$ we have that the flux is $\Phi_0$ and for $u>u_0$
$A(u)=\frac{\Phi_0}{2\pi R \sinh u}$ , so that 
\begin{equation}
\int_0^u A(u) \,du = \int_0^{u_0} A(u) \,du + \int_{u_0}^u \frac{\Phi_0}{2\pi R \sinh u} \,du\,.
\end{equation}
As a consequence, for $u>u_0$  $g_{1,0}$ given in ({\ref{g10})
will take the form
\begin{equation}
g_{1,0}(u)= 
N \left(\tanh \frac{u}2 \right)^{\lambda - \frac{\Phi_0}{\phi_0}} \,,
\end{equation}
where $\phi_0= {2\pi c \hbar}/q$ is the quantum of flux.
Hence, condition (b) of (\ref{behaviour}) will not be satisfied. In conclusion,
we see that a magnetic field with a compact support and finite flux can not
lead to a physical zero energy ground state on the hyperboloid. This is contrary to what
happens in the case of flat graphene, where the existence and degeneracy of the zero ground energy level depends on the finite flux, a property that is known as Aharonov--Casher theorem \cite{Aharonov,Kuru18}. We have seen that on the hyperboloid, only when the flux is divergent the ground state can exist, and its degeneracy will be described by some values of $\lambda$ (we will show some examples in the following section).


\section{Solvable cases of magnetic potentials}
Now, we will consider some special cases for the function $A(u)$
such that the eigenvalue equation (\ref{hsc2}) with $q=-e$ can be
solved analytically:
\begin{enumerate}
\item
\begin{equation}\label{a1}
A(u)=- \frac{c\,\hbar}{e R}\, A_0\, \coth{u}\, ,
\end{equation}
\item
\begin{equation}\label{a2}
A(u)=\frac{c\,\hbar}{e\,R}\,(-\frac{\lambda'}{\sinh{u}}+C_1\,\coth{u}-\frac{D_1}{C_1})\,
,
\end{equation}
\item
\begin{equation}\label{a3}
A(u)=\frac{c\,\hbar}{e\,R}\,(-\frac{\lambda'}{\sinh{u}}-C_2\,\tanh{u}-\frac{D_2}{C_2})\,
,
\end{equation}
\item
\begin{equation}\label{a4}
A(u)=\frac{c\,\hbar}{e\,R}\,(-\frac{\lambda'}{\sinh{u}}-C_3\,\tanh{u}-D_3\,{\rm
sech}{u})\, ,
\end{equation}
\end{enumerate}
where the parameters $\lambda'$, $D_k$ and $C_k$ are real constants. For these cases
the corresponding magnetic fields are given by (\ref{bb}).

The case $A(u)=0$, of a null magnetic
field, does not support bound states, so it will not be considered. 
The first
case (i) leads to constant magnetic field and the second case (ii) to a decaying magnetic field which give rise to bound states, the analytic solutions can
be found when the angular momentum $\lambda$ coincides with the parameter $\lambda'$
of the potential.
The third and the fourth cases lead to magnetic fields and effective potentials that in general have bad boundary conditions at the origin; only for some special values of the parameters they are acceptable and correspond to finite magnetic fields
that have a finite limit in $u\to \infty$ and in $u\to 0$.  In this section, we will deal  with the first and the second cases in detail; the last cases will be briefly commented.

\subsection{Case (i)}
The vector potential (\ref{a1}) leads to a constant magnetic field
\begin{equation}\label{b1}
{B}_{u,\varphi}(u)=-\frac{B_0}{R^2} \,,
\end{equation}
where $B_0= A_0(\frac{c\,\hbar}{e})$ is constant and
the sign determines the orientation of
the magnetic field. Therefore, this system can be considered as the Landau system on
the hyperboloid for massless relativistic particles.
Here, the vector potential gives rise to the following
superpotential
\begin{equation}\label{w1}
W(u)=A_0\,\coth{u}-\lambda\,{\rm cosech}\,{u}, \qquad A_0<\lambda\,
\end{equation}
and to the partner potentials
\begin{equation}\label{v111}
V_1(u)=A_0^2+(A_0^2+\lambda^2+A_0)\,{\rm cosech}^2{u}-\lambda
(2A_0+1)\coth{u}\,{\rm cosech}\,{u}\, ,
\end{equation}
\begin{equation}\label{v121}
 V_2(u)=A_0^2+(A_0^2+\lambda^2-A_0)\,{\rm cosech}^2{u}-\lambda
(2A_0-1)\coth{u}\,{\rm cosech}\,{u}\,  .
\end{equation}
They are shape invariant potentials \cite{Cooper}  satisfying
$V_2(u,A_0+1)=V_1(u,A_0)+2 A_0+1$.

In this case,  $g_{1,0}$ is annihilated by $L^-$, as it was shown in Sect.~4, where
\begin{equation}
L^- = \partial_u -\lambda\, {\rm cosech} u + A_0\, {\rm coth} u\,.
\end{equation} 
Thus, the zero energy ground state wavefunction is
\begin{equation}
g_{1,0}(u) = N (\tanh \frac{u} 2)^\lambda \, \frac1{(\sinh u)^{A_0}}
\end{equation}
and  its asymptotic behaviour is
\begin{equation}\label{behaviour2}
\begin{array}{lll}
(a)\quad {\rm in}\  &u \to 0,\qquad &g_{1,0}(u) \approx u^{\lambda-A_0},
\\[1.5ex]
(b)\quad {\rm in}\  &u \to \infty,\qquad &g_{1,0}(u) 
\approx\displaystyle \frac1{(\sinh u)^{A_0}}.
\end{array}
\end{equation}
This means that the ground state so defined is physically acceptable ($g_{1,0}(u) \to 0 $) if
\begin{equation}\label{al}
A_0>0\ \quad {\rm and}\quad\  \lambda-A_0\geq0\, .
\end{equation}
In these conditions the ground state has zero energy and has infinite
degeneracy determined by the (half odd) values of the total angular momentum $\lambda$
such that $\lambda-A_0\geq0$. 
Since the magnetic field is constant for all $u$, the total flux is infinite, which agree with the previous discussion.

The energy eigenvalues are given by
\begin{equation}\label{ee1}
\varepsilon_{0}=\varepsilon_{1,0}=0,\qquad
\varepsilon_{n}=\varepsilon_{1,n}=\varepsilon_{2,n-1}=A_0^2-(A_0-n)^2,\qquad
n=1,2\dots [A_0]\, ,
\end{equation}
where $[A_0]$ is the integer part of $A_0$ and all of them have the same infinite degeneracy.
The eigenfunctions have the form
\begin{eqnarray}\label{g11}
g_{1,n}(w(u))&=(w-1)^{s^-/2}\,(w+1)^{-s^+/2}
 P_n^{(s^--1/2,-s^+-1/2)}(w(u)),
\\[1ex]
g_{2,n}(w(u))&=(w-1)^{(s^--1)/2}\,(w+1)^{-(s^+-1)/2}
 P_n^{(s^-+1/2,-s^++1/2)}(w(u)),\label{g12}
\end{eqnarray}
where $s^-=\lambda-A_0$, $s^+=\lambda+A_0$ and  $P_n^{(a,b)}(w(u))$
are Jacobi polynomials, $a,b>-1$, $w(u)=\cosh{u}$ \cite{Cooper}.
These solutions are acceptable if $A_0$ satisfies the  above conditions 
(\ref{al}).

Finally,  the eigenvalues of the Dirac-Weyl Eq.~(\ref{dwsc}) are
\begin{equation}\label{evg1}
{\cal{E}}_{\pm,n}=\pm\frac{1}{R}\,\sqrt{A_0^2-(A_0-n)^2}
\end{equation}
and the eigenfunctions can be read from (\ref{png}) substituting the
functions $g_{1,n}$ and $g_{2,n-1}$ of (\ref{g11}). 
The effective potentials $V_1, V_2$ and the functions $g_{1,1},\,
g_{2,0}$ corresponding to the first excited level
are displayed in Fig.~\ref{case1}. Some eigenvalues of 
the Dirac-Weyl Hamiltonian (Landau levels on the hyperboloid) and the susy partner effective Hamiltonians can be seen in Fig.~\ref{case11}. In conclusion,
there is a finite number of Landau levels, including the zero energy level,
each one with an infinite degeneracy labeled by the values of the total momentum $\lambda$
satisfying (\ref{al}).

\begin{figure}
\centering
\includegraphics[width=0.4\textwidth]{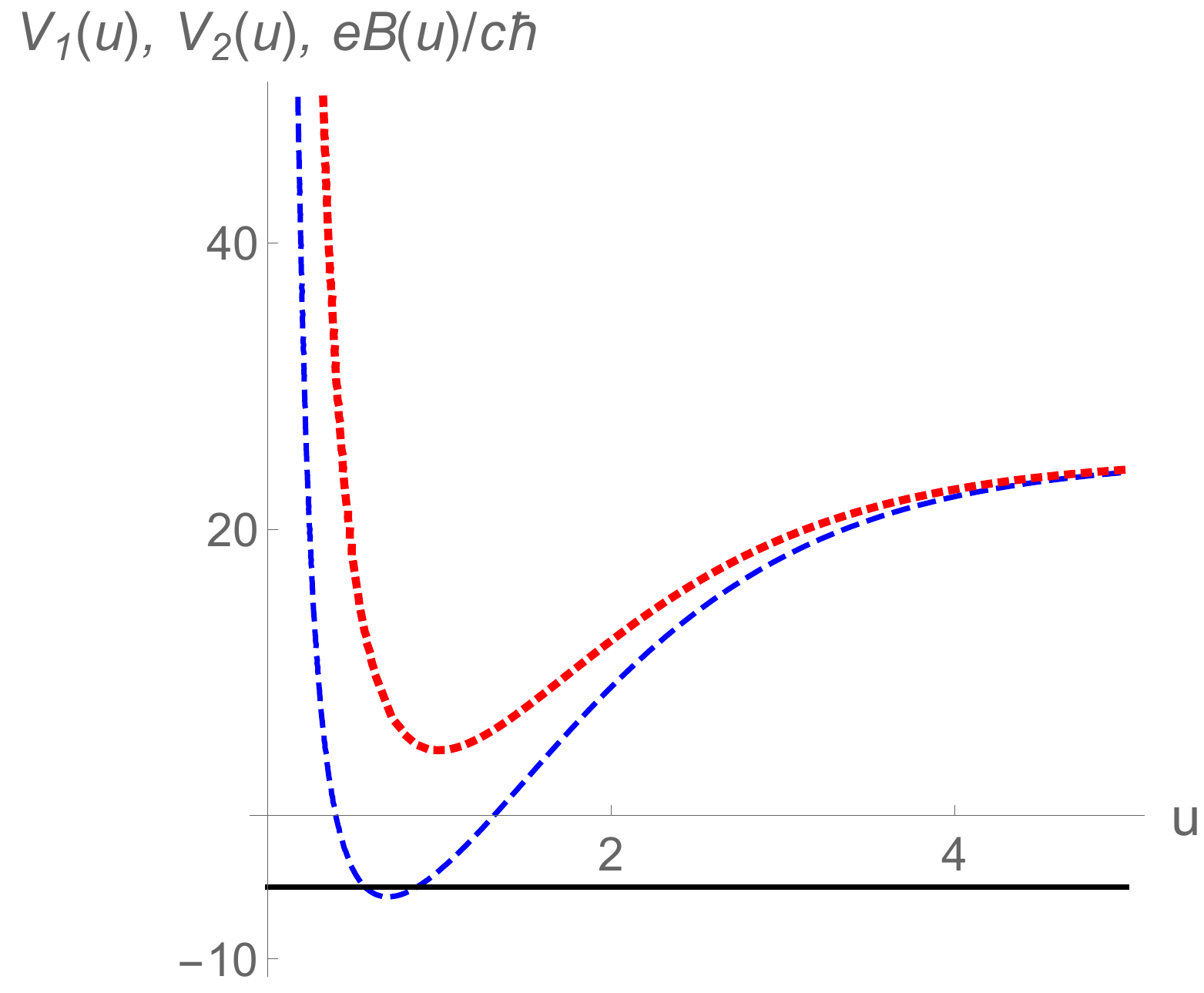}\quad
\includegraphics[width=0.4\textwidth]{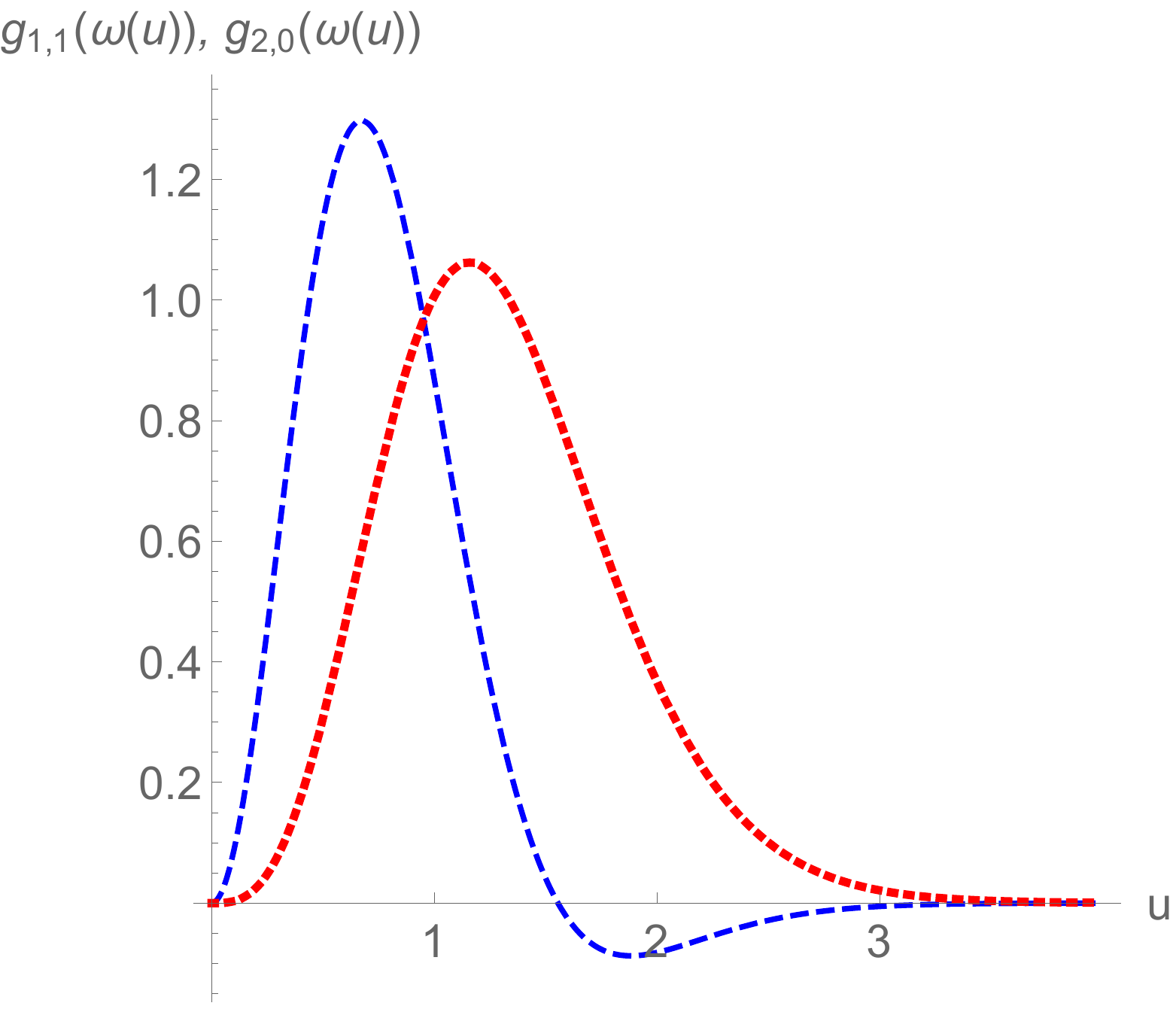}
\caption{Plot of the potentials  $V_1, V_2$ for case (i)  with
constant magnetic field (left) and the wavefunctions
$g_{1,1},\,g_{2,0}$  (right) for $A_0=5, \,\lambda=7$. Dashed lines
are for  $V_1, \,g_{1,1}$, dotted lines for $V_2,\, g_{2,0}$ and the
continuous line is for the magnetic field.\label{case1}}
\end{figure}

\begin{figure}\label{case11}
\centering
\includegraphics[width=0.4\textwidth]{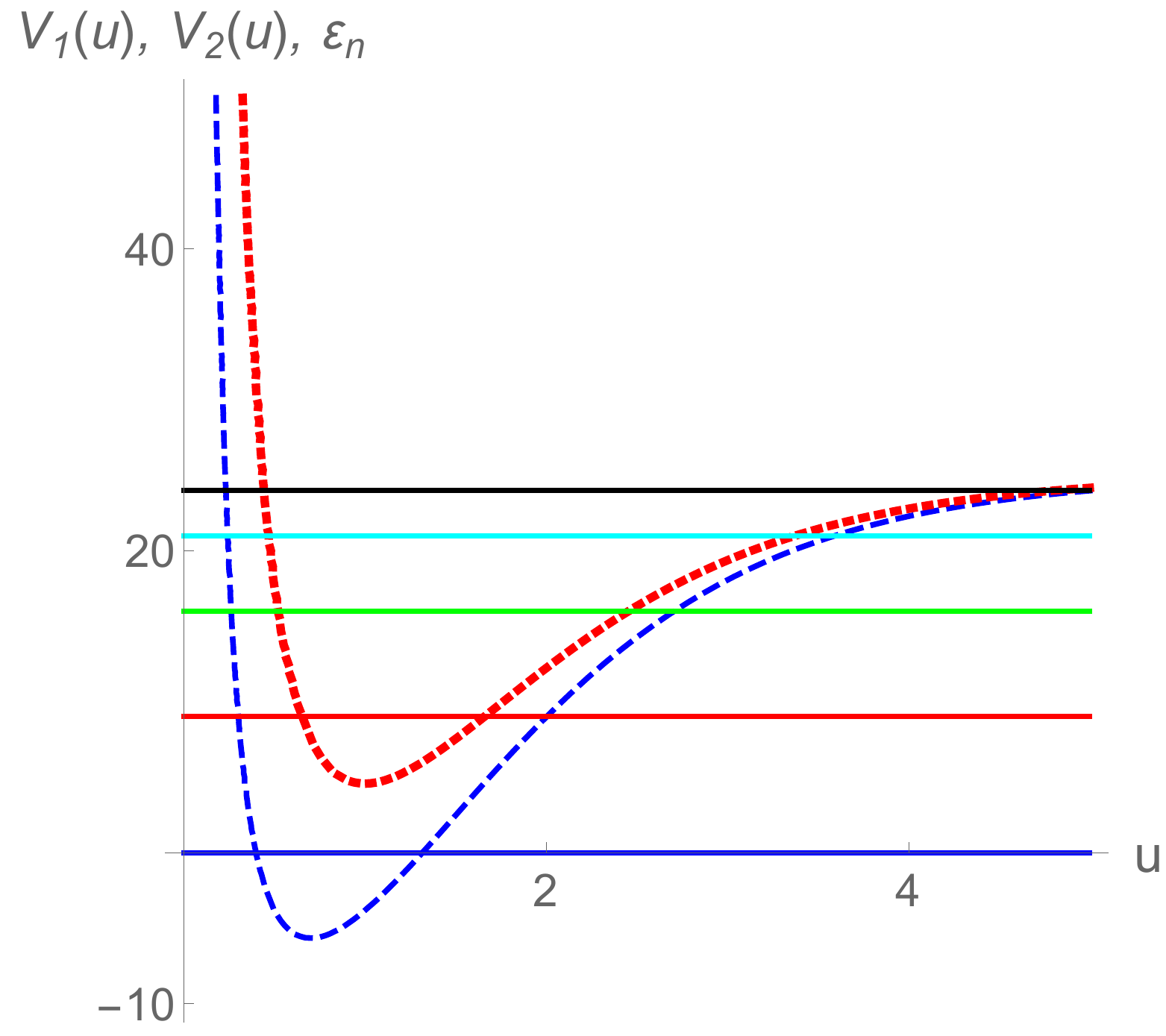}\quad
\includegraphics[width=0.4\textwidth]{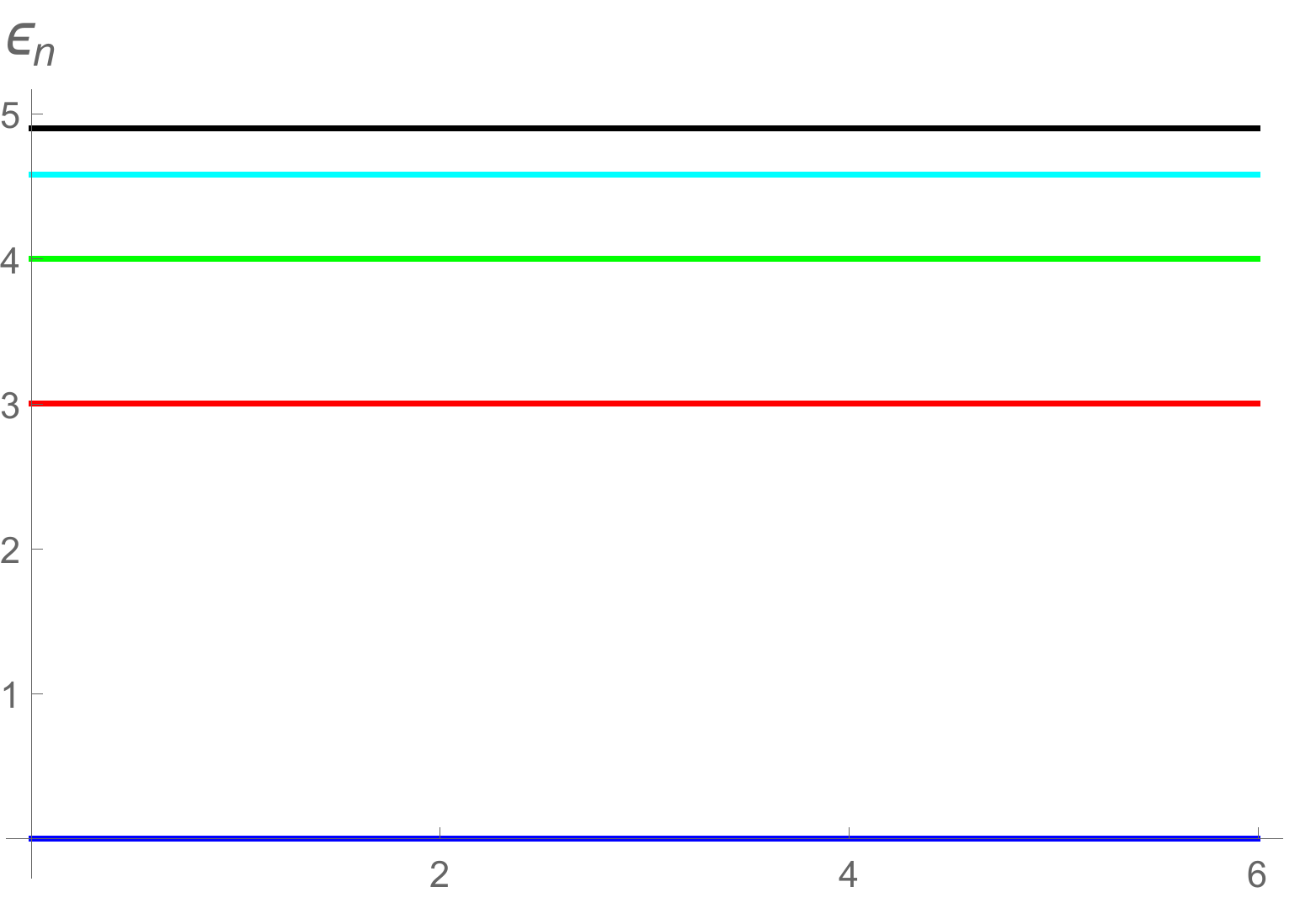}
\caption{Plot of the potentials  $V_1$ (dashed line), $V_2$  (dotted line)  and the corresponding eigenvalues $\varepsilon_{n}$ for case (i)
(left) and the eigenvalues of Dirac-Weyl Hamiltonian ${\cal{E}}_{+,n}$ (right) for $n=0$ (blue, bottom), $n=1$ (green), $n=2$ (red), $n=3$ (cyan), $n=4$ (black, top). \label{case11}}
\end{figure}


\subsection{Case (ii)}
The vector potential (\ref{a2}) 
gives the following magnetic field
\begin{equation}\label{b2}
{B}_{u,\varphi}=\frac{c\,\hbar}{e\,R}\left(\frac{C_1}{R}-\frac{D_1}{R\,C_1}\coth{u}\right)\, ,
\end{equation}
where $C_1$ and $D_1$ are real constants. This field is singular at the origin
and, as it will be shown in our discussion, it goes to a negative constant in $u\to \infty$.
If $\lambda=\lambda'$, the corresponding
superpotential has the form
\begin{equation}\label{w2}
W(u)=\frac{D_1}{C_1}-C_1\,\coth{u}\, .
\end{equation}
Then, the partner potentials obtained from (\ref{v12}) are
\begin{equation}\label{v112}
V_1(u)=\frac{D_1^2}{C_1^2}+C_1^2+C_1(C_1-1)\,{\rm
cosech}^2{u}-2\,D_1\,\coth{u}\, ,
\end{equation}
\begin{equation}\label{v122}
V_2(u)=\frac{D_1^2}{C_1^2}+C_1^2+C_1(C_1+1)\,{\rm
cosech}^2{u}-2\,D_1\,\coth{u}\, .
\end{equation}
In this case, the ground wavefunction  $g_{1,0}$ is also annihilated by $L^-$ provided
$D_1>C_1^2$. These
potentials are inside the class of Eckart potentials
\cite{Cooper}. Then,  the corresponding energy eigenvalues are given
by
\begin{equation}\label{ee2}
\fl 
\varepsilon_{0}=\varepsilon_{1,0}=0,\qquad
\varepsilon_{n}=\varepsilon_{1,n}=\varepsilon_{2,n-1}=C_1^2-(C_1+n)^2-\frac{D_1^2}{(C_1+n)^2}+\frac{D_1^2}{C_1^2},
\end{equation}
where $n=1,2\dots$.

The eigenfunctions have the form
\begin{eqnarray}\label{g112}
g_{1,n}(w(u))&=(w-1)^{s_1^{+}/2}\,(w+1)^{s_1^{-}/2}
 P_n^{(s_1^{+},s_1^{-})}(w(u)),\\[1ex]
 g_{2,n}(w(u))&=(w-1)^{s_2^{+}/2}\,(w+1)^{s_2^{-}/2}
 P_n^{(s_2^{+},s_2^{-})}(w(u)),\label{g122}
\end{eqnarray}
where $P_n^{(a,b)}(w(u))$ are Jacobi polynomials, $a,b>-1$,
$w(u)=\coth{u}$ and $s_1^{\pm}={\pm}\frac{D_1}{(C_1+n)}-(C_1+n)$,
$s_2^{\pm}={\pm}\frac{D_1}{(C_1+n+1)}-(C_1+n+1)$
\cite{Cooper}. These
solutions are acceptable if $D_1$ and $C_1$ satisfy the condition
$D_1> C_1^2$.

Therefore, the eigenvalues of the Dirac-Weyl Eq.~(\ref{dwsc}) are
\begin{equation}\label{evg1}
{\cal{E}}_{\pm,n}=\pm\frac{1}{R}\,\sqrt{C_1^2-(C_1+n)^2-\frac{D_1^2}{(C_1+n)^2}+\frac{D_1^2}{C_1^2}}
\end{equation}
and the eigenfunctions can be read from (\ref{png}) substituting the
functions $g_{1,n}$ and $g_{2,n-1}$ of (\ref{g112})-(\ref{g122}). Fig.~\ref{case2} shows
the effective potentials $V_1,\,V_2$ and the functions
$g_{1,1},\,g_{2,0}$ corresponding to the first excited level. It can be seen the eigenvalues of susy partner Hamiltonians and the Dirac-Weyl Hamiltonian in Fig.~\ref{case21}. 

Let us mention that if $\lambda'\neq \lambda$ the spectrum can not be solved
analytically, but the zero energy ground states will exist if $D_1> C_1^2$
and $\lambda\geq \lambda'$.

\begin{figure}
\centering
\includegraphics[width=0.4\textwidth]{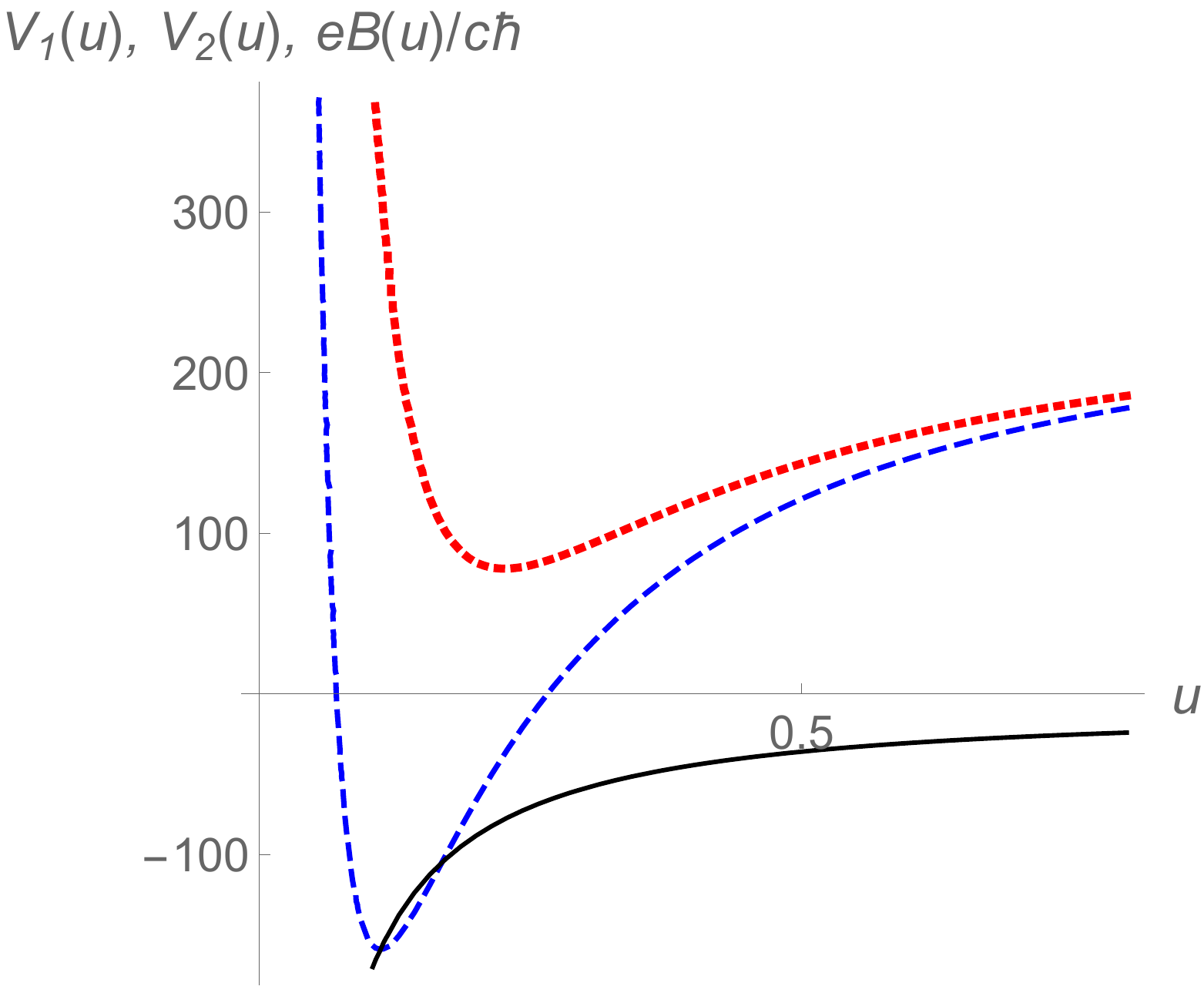}\quad
\includegraphics[width=0.4\textwidth]{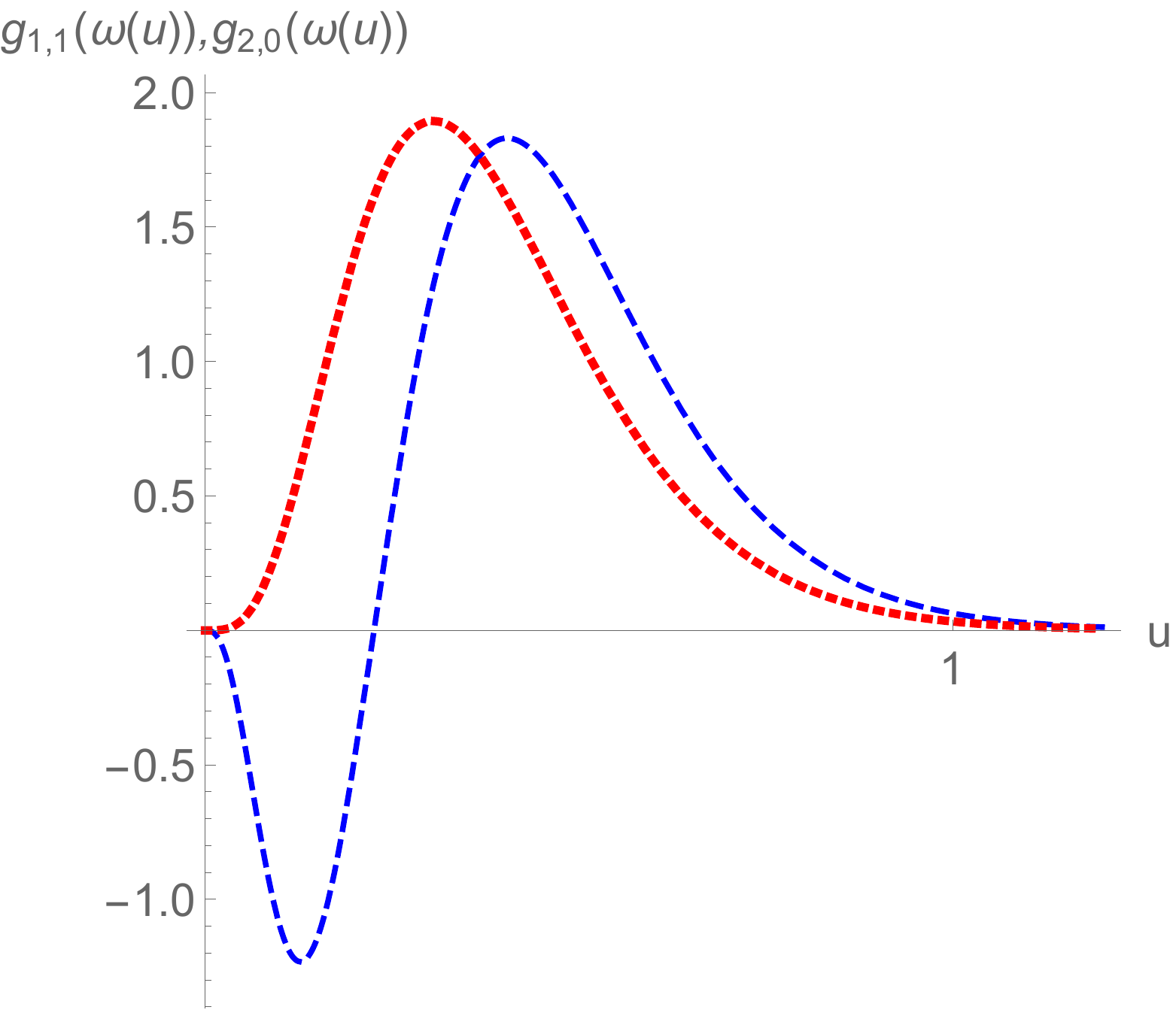}
\caption{Plot of the  Eckart potentials  $V_1, V_2$ for case (ii)
(left) and the wavefunctions $g_{1,1},\,g_{2,0}$ of the first
excited state (right) for $C_1=3, \,D_1=54$. Dashed lines are for
$V_1, \,g_{1,1}$, dotted lines for $V_2,\, g_{2,0}$ and the
continuous line is for the magnetic field.\label{case2}}
\end{figure}

\begin{figure}
\centering
\includegraphics[width=0.4\textwidth]{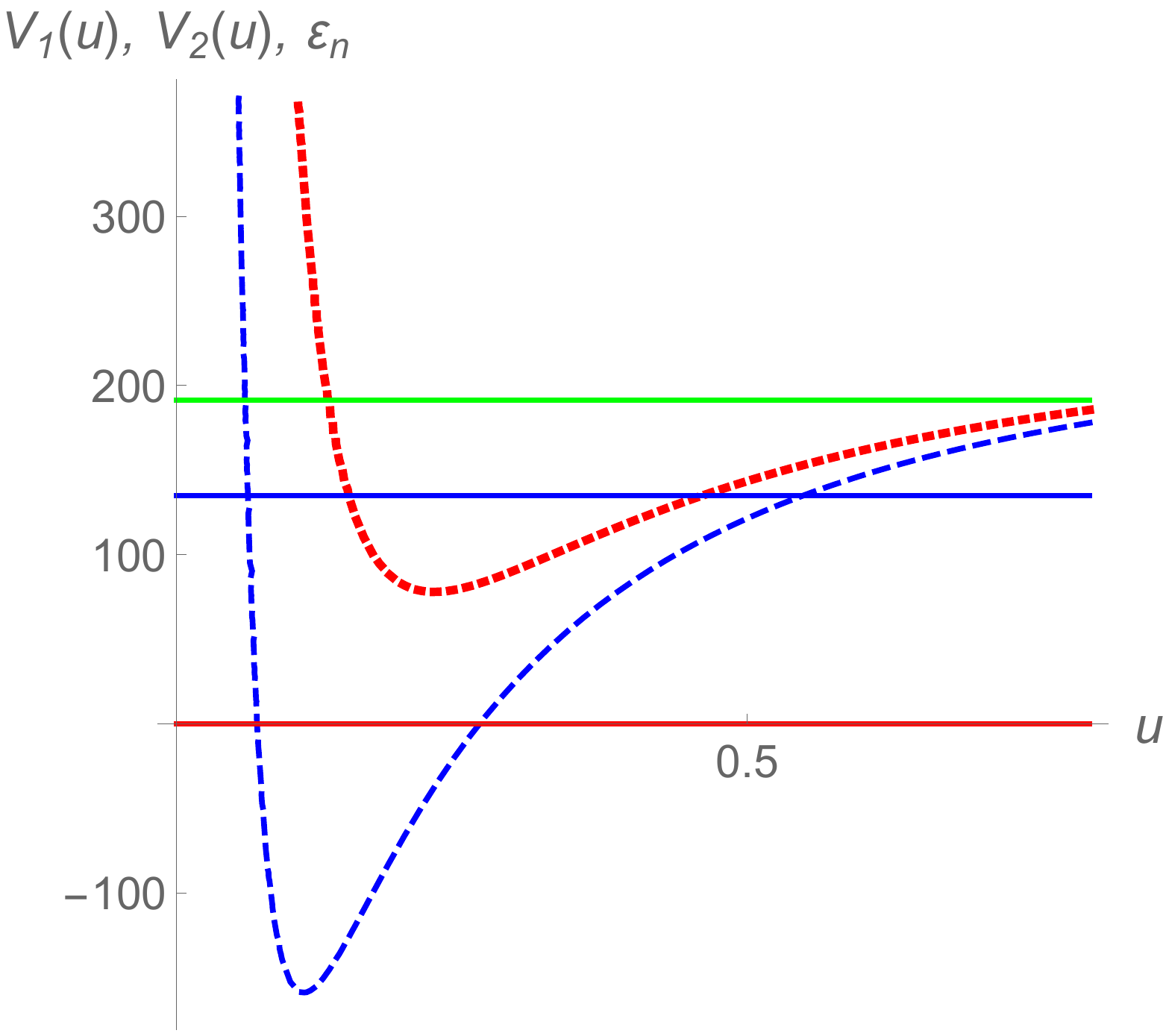}\quad
\includegraphics[width=0.4\textwidth]{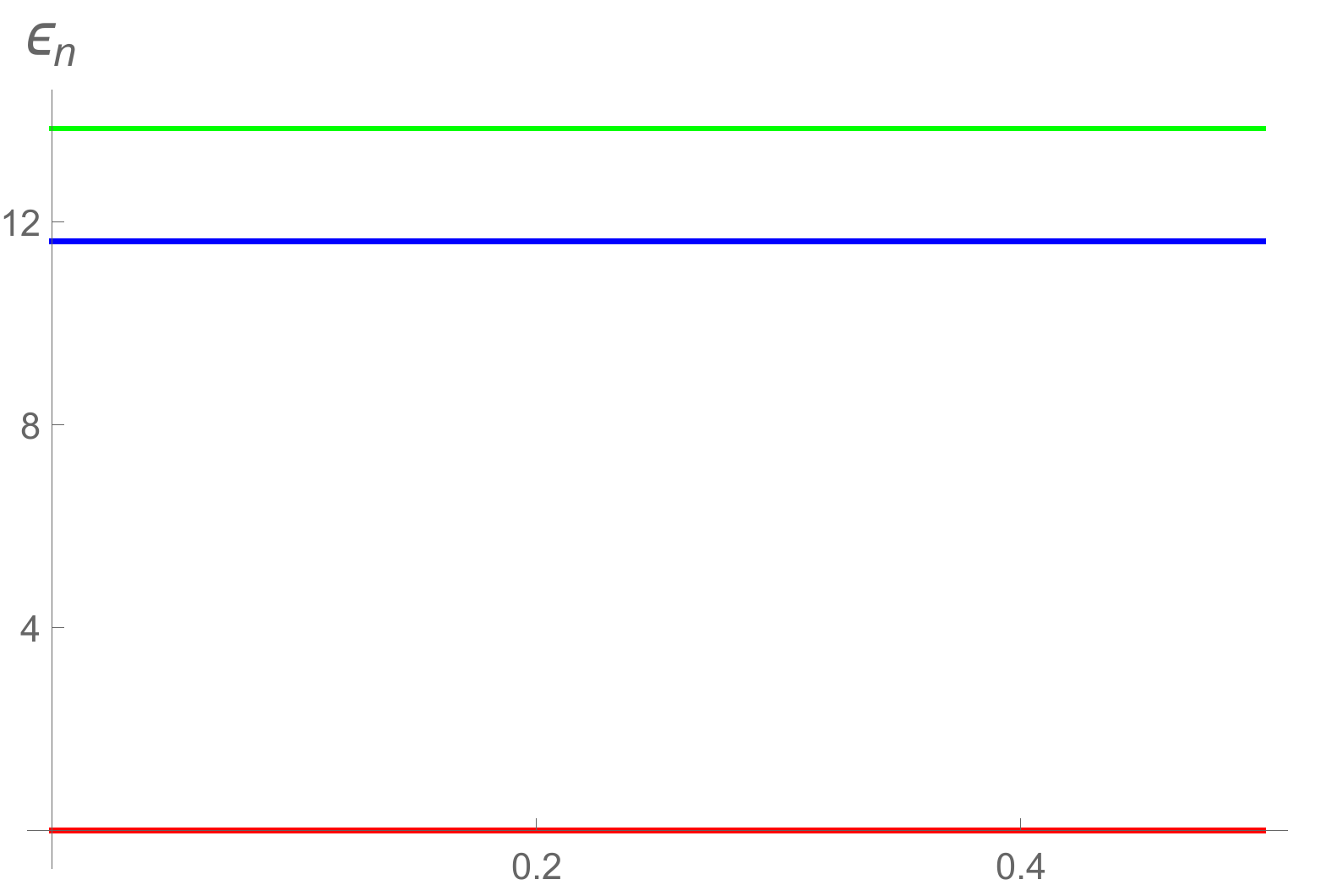}
\caption{Plot of the  Eckart potentials  $V_1$ (dashed line), $V_2$  (dotted line)  and the corresponding eigenvalues $\varepsilon_{n}$ for case (ii)
(left) and the eigenvalues of Dirac-Weyl Hamiltonian ${\cal{E}}_{\pm,n}$ (right) for $n=0$ (red), $n=1$ (blue), $n=2$ (green). \label{case21}}
\end{figure}

\subsection{Cases (iii) and (iv) }

In these two cases only when the parameters $D_2$ and $D_3$ vanish,
we will have magnetic fields with reasonable boundary conditions
at the origin. For such values the potential will be
\begin{equation}
A(u)=\frac{c\,\hbar}{e\,R}\,(-\frac{\lambda'}{\sinh{u}}-C_2\,\tanh{u})\, .
\end{equation} 
Then, the magnetic field will take the form
\begin{equation}\label{b3}
{B}_{u,\varphi}=-\frac{c\,\hbar}{e\,R}\left(\frac{C_2}{R}(1+\rm{sech}^2u) \right)\, .
\end{equation}
When $u\to\infty$, the magnetic field will tend to a constant and at the origin has a minimum. Therefore, the flux will be infinite and the zero energy ground state given by $g_{1,0}$ will be good. If $\lambda = \lambda'$ the corresponding superpotential in this case is
\begin{equation}\label{w3}
W(u)= C_2\,\tanh{u}\,,
\end{equation}
and the partner potentials found from (\ref{v12}) are P\"oschl--Teller
potentials \cite{Cooper},
\begin{equation}\label{v113}
V_1(u)= C_2^2-C_2(C_2+1)\,{\rm
sech^2}{u} \, ,
\end{equation}
\begin{equation}\label{v123}
V_2(u)= C_2^2-C_2(C_2-1)\,{\rm
sech^2}{u} \, .
\end{equation}
The eigenvalues and eigenfunctions of the Dirac-Weyl equation are
obtained as
\begin{equation}\label{ee3}
\varepsilon_{0}=\varepsilon_{1,0}=0,\qquad
\varepsilon_{n}=\varepsilon_{1,n}=\varepsilon_{2,n-1}=C_2^2-(C_2-n)^2\,,
\end{equation}
where $n=1,2\dots$\, .

\begin{figure}
\centering
\includegraphics[width=0.4\textwidth]{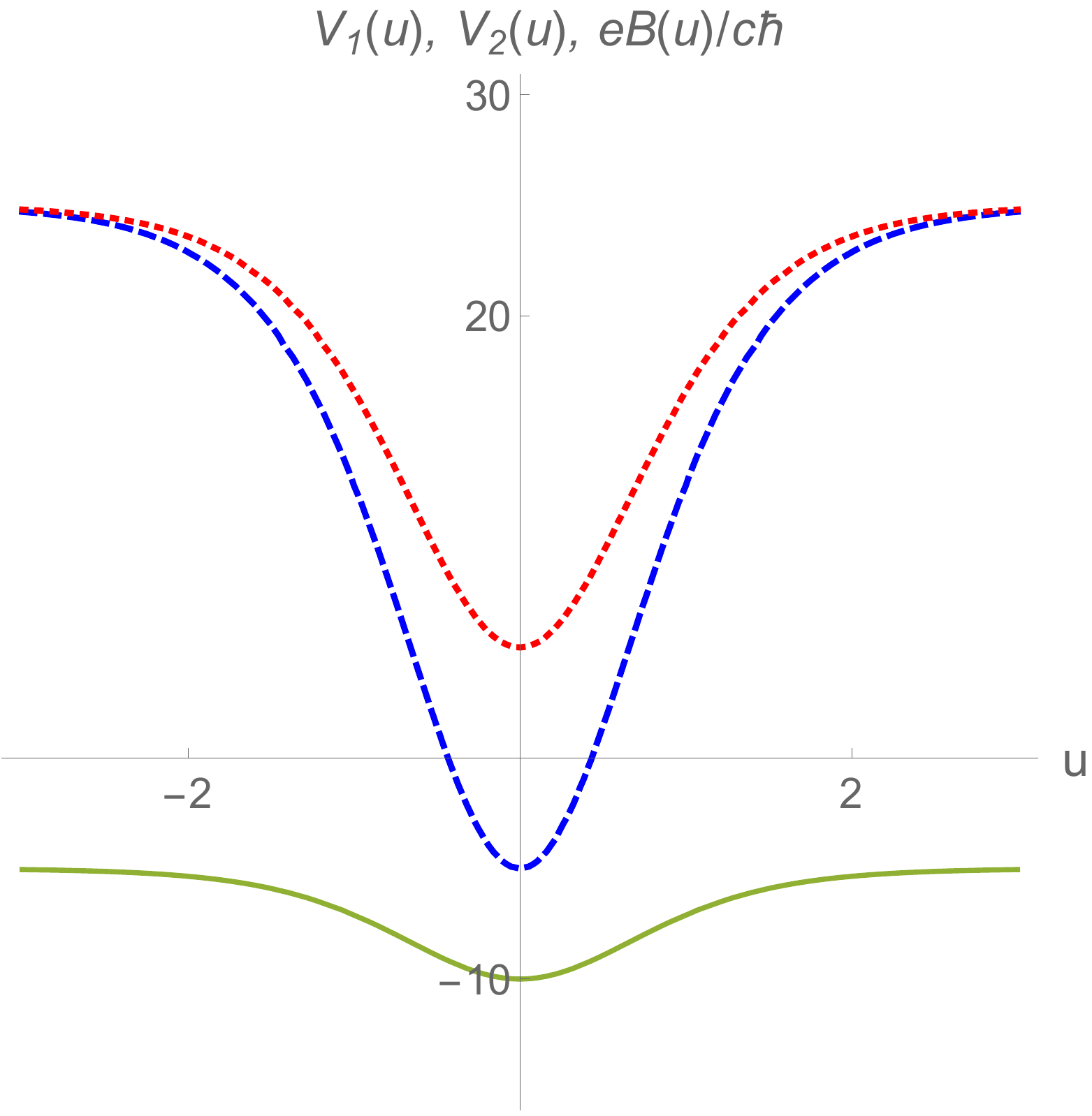}\quad
\includegraphics[width=0.4\textwidth]{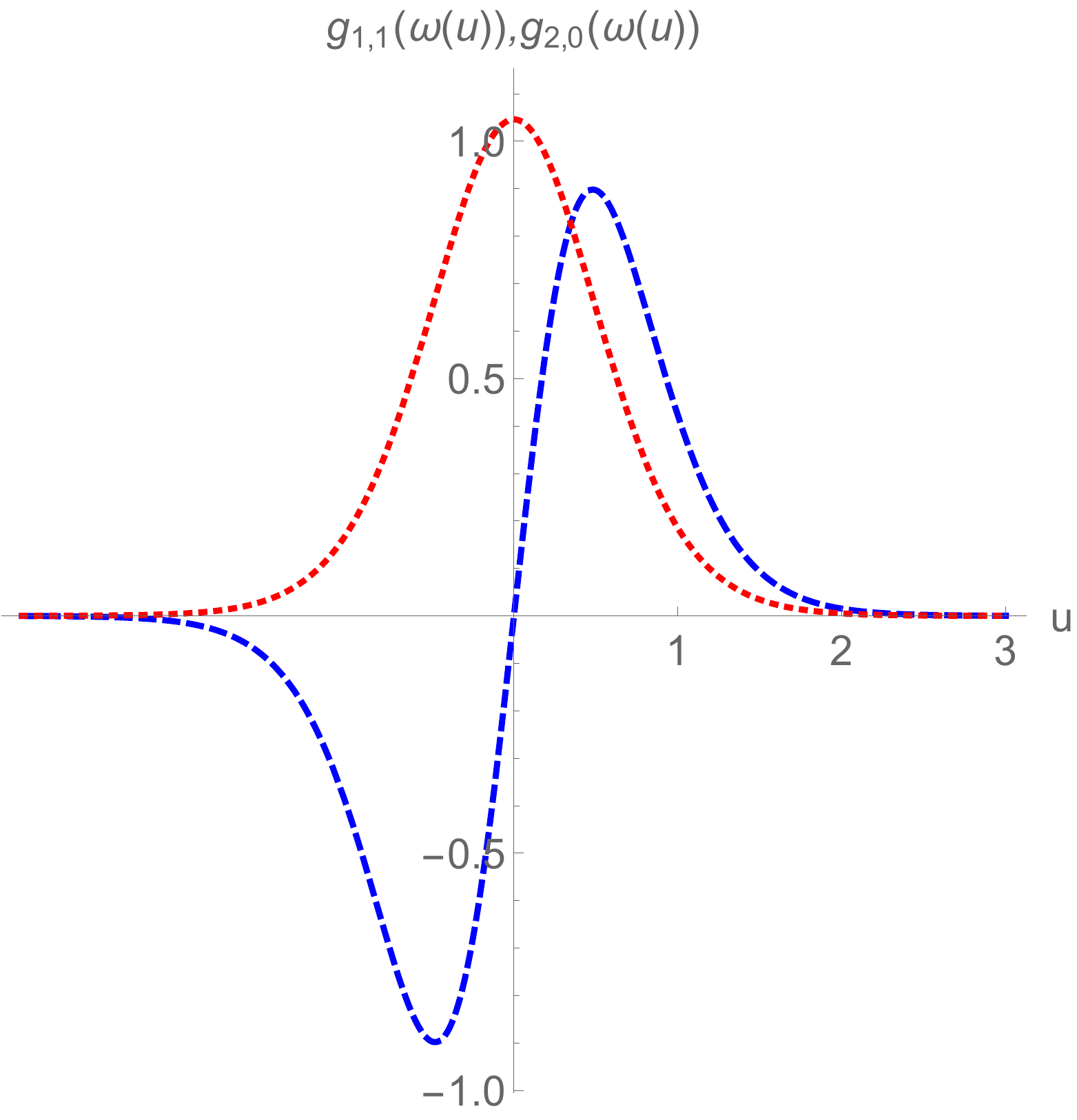}
\caption{Plot of the  P\"oschl-Teller potentials  $V_1, V_2$  for case
(ii) (left) and the wavefunctions $g_{1,1},\,g_{2,0}$ of the first
excited state (right) for $C_2=5$. Dashed lines are for
$V_1, \,g_{1,0}$, dotted lines for $V_2,\, g_{2,1}$ and the
continuous line is for the magnetic field. \label{case3}}
\end{figure}

The eigenfunctions have the form
\begin{eqnarray}\label{g113}
g_{1,n}(w(u))&=(1-w)^{s_1/2}\,(1+w)^{s_1/2}
 P_n^{(s_1,s_1)}(w(u)),\\[1ex]
 g_{2,n}(w(u))&=(1-w)^{s_2/2}\,(1+w)^{s_2/2}
 P_n^{(s_2,s_2)}(w(u)),\label{g123}
\end{eqnarray}
where $P_n^{(a,b)}(w(u))$ are Jacobi polynomials, $a,b>-1$,
$w(u)=\tanh{u}$ and $s_1 = (C_2-n)$,
$s_2 = (C_2-n-1)$  \cite{Cooper}. These
solutions are acceptable if  $C_2>0$.

Therefore, the eigenvalues of the Dirac-Weyl Eq.~(\ref{dwsc}) are
${\cal{E}}_{\pm,n}=\pm\frac{1}{R}\,\sqrt{\varepsilon_{n}}$ and the
eigenfunctions can be read from (\ref{png}) substituting the
functions $g_{1,n}$ and $g_{2,n-1}$ of (\ref{g113})-(\ref{g123}). Fig.~\ref{case3} displays
the effective potentials $V_1,\,V_2$ and the functions
$g_{1,1},\,g_{2,0}$ corresponding to the first excited level in the real line.

\section{Conclusions}

In this paper, we have studied the system of a massless charged particle
on a hyperbolic surface under a rotationally symmetric perpendicular
magnetic field. This problem can be identified with that of $\pi$--electrons
in a deformed graphene sheet having this shape.

Instead of defining the Dirac--Weyl equation on this surface through
the metric and spin connection, we have preferred to do it by formulating
the Dirac--Weyl equation in an appropriate ambient space and then,
restrict it to the hyperbolic surface. In this way, we preserve the 
rotational symmetry explicitly. This process is carried out in a straightforward
way by means of the definition of momentum operators tangent to the surface.

After making use of the rotational symmetry we have arrived to a reduced
Dirac--Weyl equation for two--component spinors in the `radial' variable $u$,
that displays the minimal coupling with the magnetic potential.

One of the points that we addressed was  whether the known Aharonov-Casher theorem
on the existence and degeneracy of the ground (zero-energy) state applies in this situation. 
We have shown that for radial symmetric magnetic fields with compact support
and finite magnetic flux such zero energy modes don't exist. Indeed,  we
have considered a few analytically solvable cases 
where  there exist only a finite number of discrete
energy levels (besides the continuum spectrum). In these cases the magnetic
potential is singular at the origin, such singularity can be compensated
by the angular momentum, while the behaviour of the potential far from
the origin allows for bound states (see this behaviour in formula (\ref{behaviour2})). In other words, the mechanism for
confining  massless particles 
is quite different in the hyperbolic surface than in a flat surface.

Among the analytically  solvable cases here studied it is included the constant magnetic
field. The discrete spectrum consist of the zero energy ground level plus a finite 
number of excited levels.
The infinite degeneracy of each energy level is characterized by 
the total angular momentum that regularize the singularity of the potential at the origin.
In the other cases the Dirac-Weyl equation is solvable for just one angular
momentum ($\lambda=\lambda'$), but the conditions for the existence of zero energy ground state are the same. We should mention that the class of solvable potentials can be extended to other supersymmetric partner potentials by means of Darboux transformations.

We hope that this problem can help in different applications of graphene
surfaces deformed in a variety of shapes, as it was mentioned in the introduction.
In particular we want to see the influence of pseudo magnetic fields on
the energy levels and on the degeneracy of the ground level.

 \section*{Acknowledgments}
This work is partially supported by Junta de Castilla Le\'on, Spain (BU229P18 and VA137G18).
 \c{S}.~Kuru acknowledges the warm hospitality at Department of
Theoretical Physics, University of Valladolid, Spain.



\end{document}